\documentclass[10pt,twocolumn]{paper}
\usepackage{epsfig, graphicx}
\usepackage[english]{babel}
\usepackage[margin=1.5cm]{geometry}

\title{\center \rm \bf Solid-Liquid Phase Transition in the Octadecanoic Acid Film Adsorbed on the Toluene-Water Interface}

\author{\small \rm Aleksey M. Tikhonov$^{a,b}$\/\thanks{tikhonov@kapitza.ras.ru}\\
\small
$^a$Kapitza Institute for Physical Problems, Russian Academy of Sciences, Moscow, 119334 Russia\\
\small
$^b$Institute of Solid State Physics, Russian Academy of Sciences, Chernogolovka, Moscow region, 142432 Russia
}

\begin{document}
\maketitle

\abstract{ \it \normalsize
The structure of the soluble protonated (pH=2) octadecanoic acid film adsorbed on the saturated
hydrocarbon (n-hexane) - water and aromatic hydrocarbon (toluene)-water interfaces is studied by X-ray
reflectometry using synchrotron radiation. The experimental data demonstrate that a solid phase of a Gibbs
monolayer $26 \pm 1$\,{\AA} thick, in which aliphatic tails are perpendicular to the surface and the area per molecule is $A=18\pm 2$\,\AA$^2$, forms in the film at the n-hexane - water interface. The solid monolayer on the toluene - water interface in the adsorbed film melts when temperature increases, and this transition is caused by disordering the hydrocarbon tails of the acid. During the solid - liquid transition, the Gibbs monolayer thickness remains almost the same, $22 \pm 1$\,{\AA}. In the solid phase, we have $A=20\pm 2$\,\AA$^2$ and the angle of deviation of the molecular tails from the normal to the surface is about $\approx 30^\circ$. The density of the liquid monolayer phase with $A=24 \pm 2$\,\AA$^2$ corresponds to liquid n-octadecane.
}

\vspace{0.25in}
\normalsize

{\bf INTRODUCTION}

Thermotropic phase transitions between surface mesophases are observed in the soluble amphiphylic
substance film adsorbed on the nonpolar organic solvent (oil)-water interface. These transitions can be
both extended in temperature and characterized by sharp changes in the state of surface. The works on
studying such surface phenomena can be conventionally divided into the following two types. The works of
the first type investigate the structure of the internal interfaces in the material volume that appear due to the microscopic separation of phases with the formation of micelle and liposome solutions or microemulsions [1]. The works of the second type deal with the interfaces between macroscopically large oil and water volumes [2-6]. The authors of [7, 8] were the first to demonstrate the possibility of application of X-ray reflectometry using synchrotron radiation to determine the molecular ordering on the macroscopically flat saturated hydrocarbon (n-hexane)–water interface. Later [9-12], we used this technique to study the thermotropic phase transitions at this interface in adsorbed fatty alcohol and acid layers. The purpose of this work is to investigate the solid - liquid phase transition in the soluble octadecanoic acid film adsorbed on the aromatic hydrocarbon (toluene) - water interface by X-ray reflectometry (see Fig. 1). This interface is considered as a model interface to study, e.g., the adsorption of the high-molecular-weight oil components (asphaltens) that do not dissolve in saturated hydrocarbons [13].

\begin{figure}
\vspace{0.5in}
\hspace{0.5in}
\epsfig{file=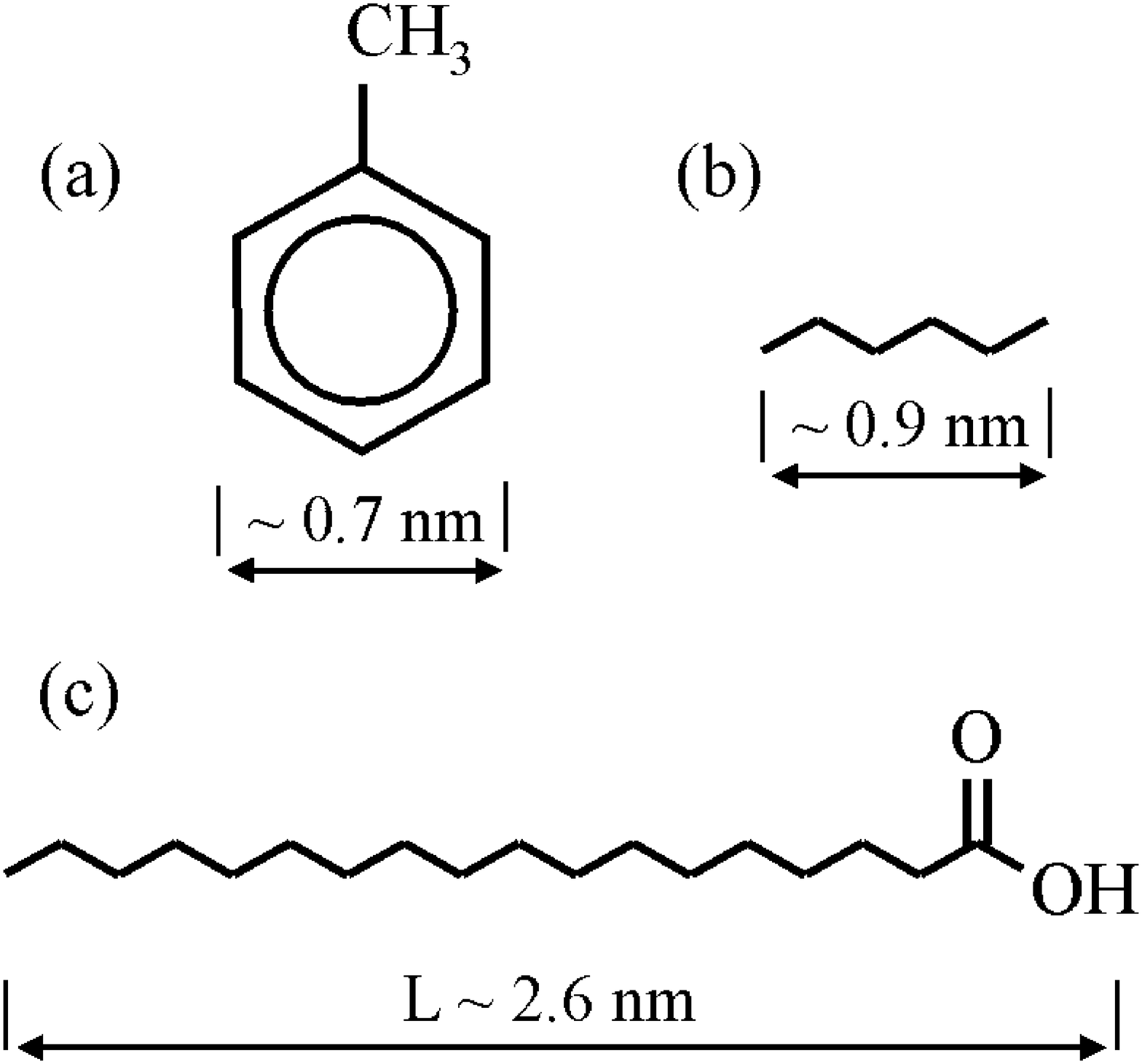, width=0.35\textwidth}

\vspace{0.15in}
\small {\bf Figure 1.} \it Molecular structures of (a) toluene C$_7$H$_8$, (b) n-hexane
C$_6$H$_{14}$, and (c) octadecanoic acid C$_{18}$H$_{36}$O$_2$.

\end{figure}

\vspace{0.25in}
{\bf EXPERIMENTAL}

All chemical components for experiments were bought at Sigma-Aldrich. Saturated hydrocarbon n-hexane
(C$_6$H$_{14}$, the density at 298\,K is $\approx 0.65$\,g/cm$^3$, the boiling temperature is $T_b\approx 342$\,K) and aromatic hydrocarbon toluene (C$_7$H$_{8}$, the density at 298\,K is $\approx 0.87$\,g/cm$^3$, the boiling temperature is $T_b\approx 384$\,Ê) were preliminarily cleaned by multiple filtration in a chromatographic column [14]. Octadecanoic acid C$_{17}$H$_{35}$COOH (stearic acid, or C$_{18}$-acid) is a monocarboxylic acid of the aliphatic series, does not dissolve in water, and is well dissolved in toluene and n-hexane. This acid was purified by recrystallization from a supersaturated solution in n-hexane at room temperature [12, 15].

The samples of the flat toluene - water (n-hexane - water) interface, which was oriented by the gravitational force, was studied in a stainless steel temperature controlled cell according to the technique from [16]. The surface tension of the interface $\gamma(T)$ in Fig. 2 was measured by the Wilhelmy plate method [17]. A solution of sulfuric acid (pH = 2) in $\sim 100$\,mL of deionized water (Barnstead, NanoPureUV) was used as the lower phase. The upper phases consisted of a $\sim50$\,mL solution of octadecanoic acid in toluene (n-hexane) with a volume concentration $ñ \approx 46$\,mmol/kg ($\approx 4.2\cdot10^{-3}$). Before being placed in the cell, these fluids were subjected to degassing in an ultrasonic bath. When reflection coefficient $R$ was measured, a sample was
"annealed": the fluid temperatures in the cell was increased to $\sim 330$\,K and was then decreased to the chosen temperature, and the sample was brought in equilibrium in several hours when the lower phase was
accurately mechanically stirred [18, 19].

C$_{18}$H$_{36}$O$_2$ acid molecules from the solution in the hydrocarbon solvent are adsorbed onto the toluene - water interface, which significantly decreases its energy. As follows from Fig. 2, a phase transition takes place in the monolayer on the interface when temperature $T$ increases (at a pressure 
$p = 1$\,atm). The phase-transition temperature ($T_c\approx 319$\,K) is determined by the C$_{18}$-acid concentration $c$ in the solvent volume, which serves as a reservoir for surfactant molecules. The change in the slope of $\gamma(T)$ is related to the relatively small change in the surface enthalpy during the transition, $\Delta H = - T_c\Delta(\partial \gamma/\partial T)_{p,c}$ $=0.03\pm 0.01$\,J/m$^2$. Note that the octadecanoic acid film adsorbed on the n-heaxane-water interface exhibits no specific features in the behavior of surface tension at p = 1\,atm in wide concentration ($10-100$\,mmol/kg)
and temperature ($290-330$\,K) ranges.

The transverse structure of the toluene-water (n-heaxane - water) interface was studied by X-ray reflectometry on the X19C station of the NSLS synchrotron [20]. In experiments, we used a focused monochromatic beam with an intensity of about $\approx 10^{11}$\,photons/s and a photon energy ($\lambda=0.825 \pm 0.002$ \,\AA). The design of the X19C station spectrometer makes it possible to
investigate the surfaces of solids, liquids, and liquid-liquid interfaces [21-26].

Figure 3 shows the kinematics of surface scattering by the interface. In the reflectometry experiment, we
have $\alpha=\beta$, where $\alpha$ is the grazing angle and $\beta$ is the angle between the surface plane and the direction to the point detector in the plane of incidence $yz$. Here, X-rays pass through the oil phase and are specularly reflected by the structure formed by the surfactant on the interface. If {\bf k}$_{\rm in}$ and {\bf k}$_{\rm sc}$ are the wave vectors of the incident and reflected beam in the detector direction, respectively, scattering vector {\bf q = k$_{\rm in}$ {\rm -} k$_{\rm sc}$} in this experiment is normal to the surface along axis $z$ opposite to the gravitational force. When reflection coefficient $R$ is measured as a function of $q_z=(4\pi/\lambda)\sin\alpha$, it is averaged over a large illumination surface area ($\sim 0.5$\,cm$^2$) because of the height of the incident beam ($>5$\,$\mu$\,m) in the $yz$ plane and the width ($\sim 2$\,mm) in the interface plane.

\begin{figure}
\hspace{0.05in}
\epsfig{file=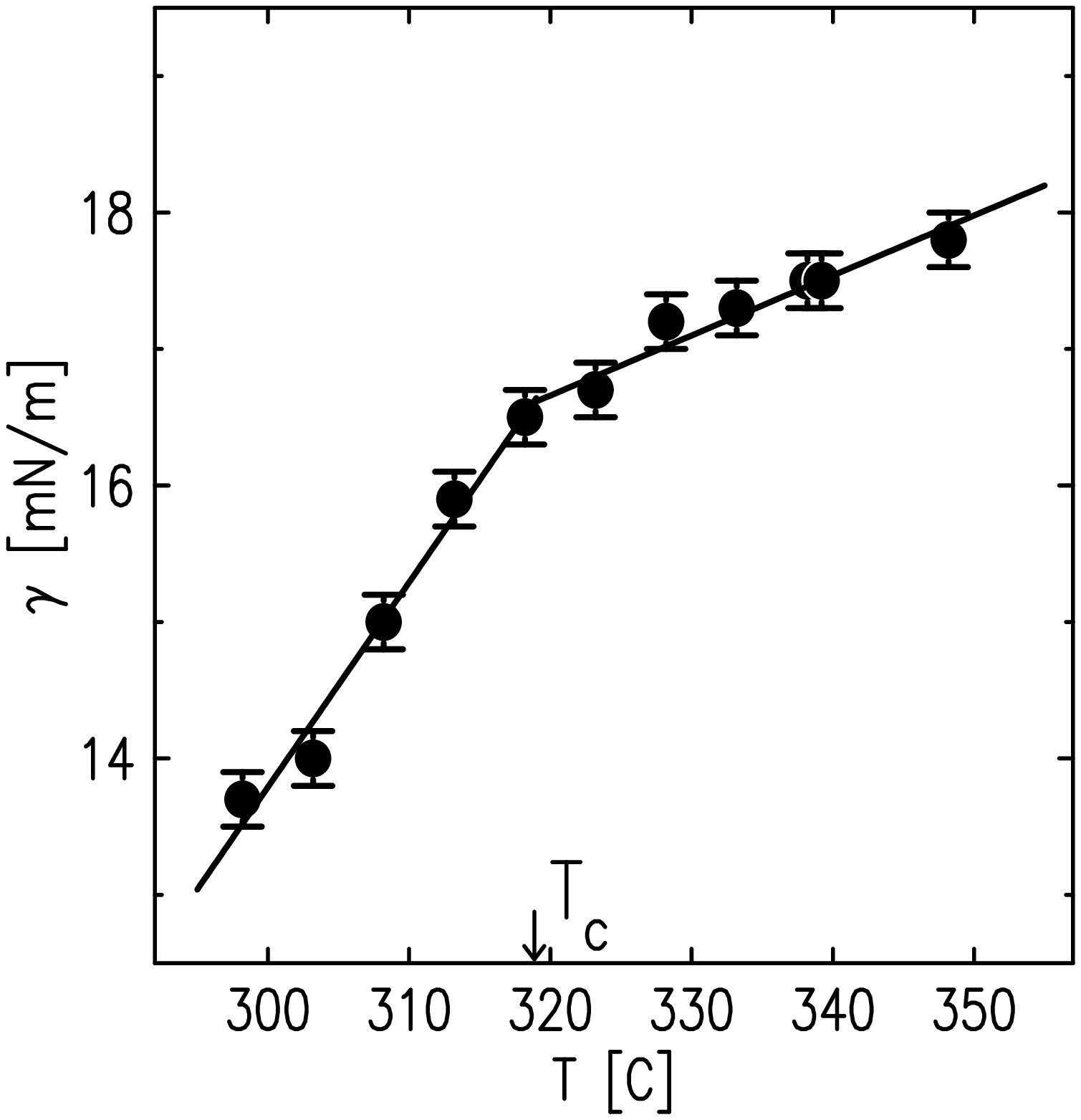, width=0.45\textwidth}

\small {\bf Figure 2.} \it Temperature dependence of the interfacial tension of the toluene-water interface at an octadecanoic acid concentration $c \approx 46$\,mmol/kg in the hydrocarbon solvent. The lines are drawn by eye and the inflection point corresponds to $T_c \approx  319$\,K.

\end{figure}

At the grazing angle lower than $\alpha_c\approx\lambda\sqrt{r_e\Delta\rho/\pi}$ (where $r_e = 2.814\cdot10^{-5}$\,{\AA} is the classic electron radius and $\Delta\rho$ is the difference between the volume electron concentrations of the fluids), the incident beam undergoes total external reflection ($R\approx 1$). Under normal conditions, the electron density is $\rho_w\approx 0.333$\,{\it e$^-$/}{\AA}$^3$ ({\it e$^-$} (e$^-$ is the electron charge) in water, $\rho_h \approx 0.69 \rho_w$ in n-hexane, and $\rho_{t} \approx 0.86 \rho_w$ in toluene. Thus, we have $\alpha_c\approx 10^{-3}$\,rad ($\approx 0.06$\,deg) for the n-hexane - water interface and $\alpha_c\approx 6 \cdot10^{-4}$\,rad ($\approx 0.03$\,deg), which is significantly lower, for the toluene - water system.

which is significantly lower, for the toluene–water system. Figure 4 shows the experimental dependences of
reflection coefficient $R$ on $q_z$ normalized by the Fresnel function
\begin{equation}
R_F(q_z)=\left(\frac{q_z-\sqrt{q_z^2-q_c^2}}{q_z+\sqrt{q_z^2-q_c^2}}\right)^2,
\end{equation}
where $q_c=(4\pi/\lambda)\sin\alpha_c$. The triangles correspond to the values of $R(q_z)/R_F(q_z)$ for the n-hexane-water interface at $T=295$\,K, and the circles and the squares,
to the values for the toluene–water interface at $T=308$
and $T=328$\,K (below and above $T_c$), respectively.

\vspace{0.25in}
{\bf THEORY}

From the experimental data $R(q_z)$, we restored the electron concentration distribution $ñ(z)$ along the
normal to the surface using the qualitative one-layer model based on the error function [27–29] (Fig. 5)
\begin{equation}
\begin{array}{c}
\displaystyle
\rho(z)=\frac{1}{2}(\rho_{0}+\rho_{2})+\frac{1}{2}(\rho_{1}-\rho_{0}){\rm erf}\left(\frac{z}{\sigma\sqrt{2}}\right)
\\ \\
\displaystyle
+\frac{1}{2}(\rho_{2}-\rho_{1}){\rm erf}\left(\frac{z+z_1}{\sigma\sqrt{2}}\right),
\\ \\
\displaystyle
{\rm erf}(x)=\frac{2}{\sqrt{\pi}}\int_0^x\exp(-y^2)dy,
\end{array}
\end{equation}
where $\rho_0$ and $\rho_2$  are the electron concentrations in water and toluene (n-hexane), respectively; $\rho_1$ is the electron concentration in the Gibbs monolayer; $z_0=0$; $z_1$ is the monolayer thickness; and ó is the root-mean-square deviation of the positions of the interfaces from their nominal values $z_0$ and $z_1$, which was taken to be equal to the "capillary width" in the calculations. This width is determined by the surface spatial frequency range covered in the experiment [30–33],
\begin{equation}
\sigma^2 \approx  \frac{k_BT}{2\pi\gamma} \ln\left(\frac{Q_{max}}{Q_{min}}\right),
\end{equation}
where $Q_{max} = 2\pi/a$ is the short-wavelength spectral limit ($a\approx 10$ {\AA} is the intermolecular distance on the order of magnitude), $Q_{min}=q_z^{max}\Delta\beta/2$ is the long wavelength limit, $\Delta\beta$$\approx 4\cdot10^{-4}$\,rad is the angular resolution of the detector in the experiment, and $q_z^{max} \approx 0.25$ {\AA}$^{-1}$. Under the experimental conditions, estimation (3) gives $\sigma = 4.0\pm0.2$\,{\AA} ($\gamma \approx 35$\,mN/m) for the n-hexane - water interface and $\sigma = 5.6\pm0.2$\,{\AA} and $\sigma = 6.1\pm0.2$\,{\AA} at $T=308$ and $T=308$\,K, respectively, for the toluene - water interface.

Profile (2) corresponds to the reflection coefficient [34, 35]
\begin{equation}
\begin{array}{l}
\displaystyle \frac{R(q_z)}{R_F(q_z)} \approx
\\ \\
\displaystyle
\left| \frac{1}{\Delta\rho}\sum_{j=0}^{1}{(\rho_{j+1}-\rho_j) e^{-iq_zz_j}} \right|^2 e^{-\sigma^2q_z\sqrt{q_z^2-q_c^2}}.
\end{array}
\end{equation}

\begin{figure}
\vspace{0.5in}
\hspace{0.1in}
\epsfig{file=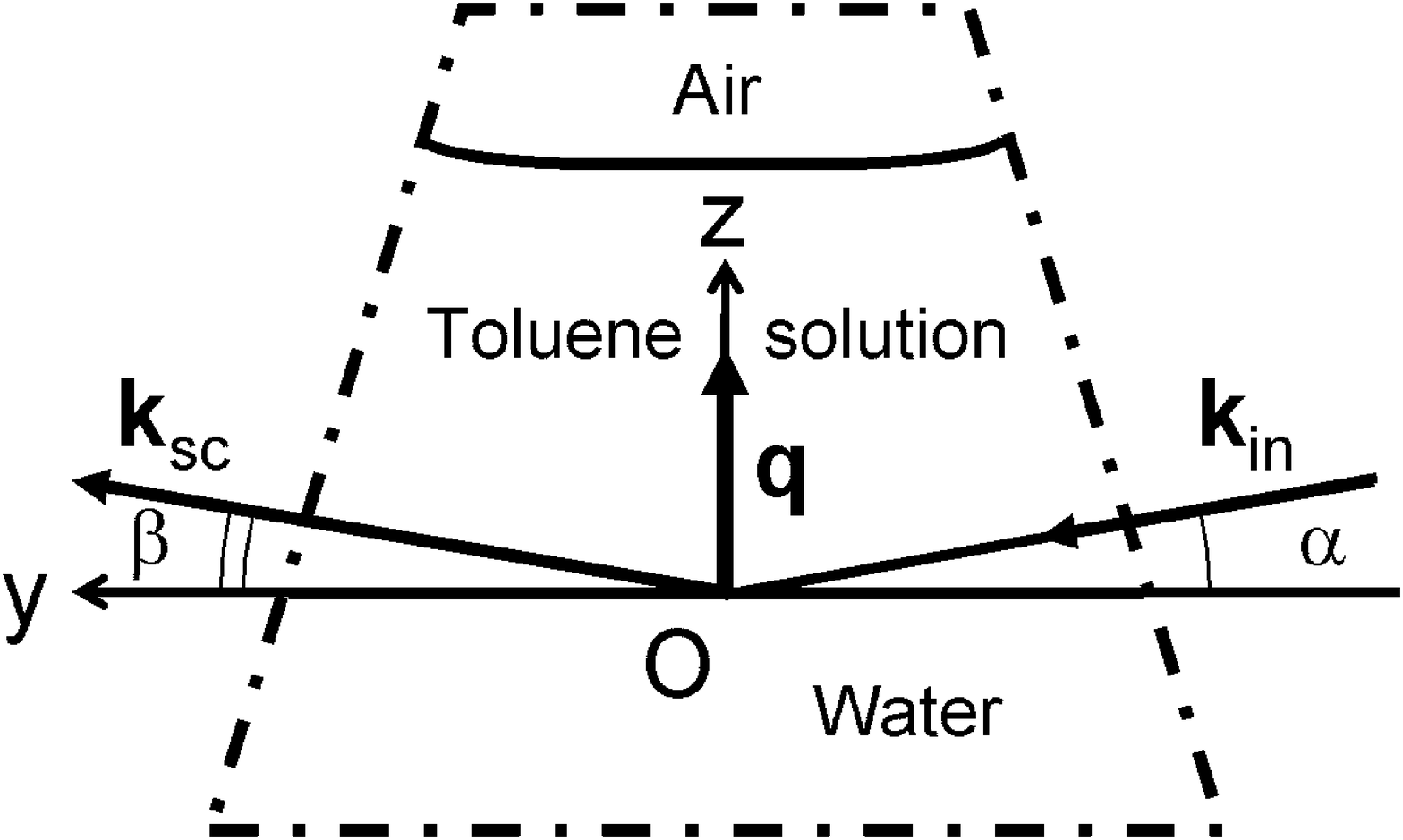, width=0.45\textwidth}

\vspace{0.15in}
\small {\bf Figure 3.} \it Kinematics of X-ray surface scattering by the toluene - water interface. $\alpha=\beta$ in the reflectometry experiment.
\end{figure}

\vspace{0.25in}
{\bf RESULTS AND DISCUSSION}

The solid lines in Fig. 4 demonstrate that the experimental $R/R_F$ dependences are well described by
Eq. (4), which has two adjustable parameters, namely, monolayer thickness $z_1$ and electron concentration in it $\rho_1$. The electron concentration profile of the adsorbed layer $\delta\rho(z)$ is obtained from Eq. (1) by the subtraction of the contributions of the bulk phases to $\rho(z)$,
\begin{equation}
\begin{array}{l}
\displaystyle \delta\rho(z)=
\rho(z) - \frac{1}{2}\rho_{0}\left[1-{\rm erf}\left(\frac{z}{\sigma\sqrt{2}}\right)\right]
\\ \\
\displaystyle
-\frac{1}{2}\rho_{2}\left[1+{\rm erf}\left(\frac{z+z_1}{\sigma\sqrt{2}}\right)\right].
\end{array}
\end{equation}
The $\delta\rho(z)$ profiles normalized by $\rho_w$ are shown in Fig. 6.

The thermodynamic properties of the soluble adsorbed film, which is considered to be a monolayer
in a first approximation (Gibbs monolayer), are described by parameters $(p, T, c)$ [2, 36-38]. In the
interface plane, this film can be both isotropic and anisotropic despite the isotropy of the bulk phases
[39]. In this system, first-order phase transitions are formally prohibited, since the formation of an equilibrium spatially inhomogeneous structure, in which the domains of two homogeneous phases coexist, from
adsorbed molecules is thermodynamically favorable in a certain vicinity of $T_c$ [40]. Both phases tend toward intermixing, since the formation of one-dimensional interfaces leads to a significant decrease in the energy [41]. As follows from the lyotropic and thermotropic phase transitions between the bulk mesophases in aqueous fatty acid solutions, one of the parameters that determine the thermodynamic state of a system is
solution pH, which affects the degree of ionization of the –COOH group [36]. Since the hydroxyl groups in
the monolayer are not ionized at pH=2 according to $^1$H proton NMR data, it is conventionally called "protonated" [42].

\begin{figure}
\epsfig{file=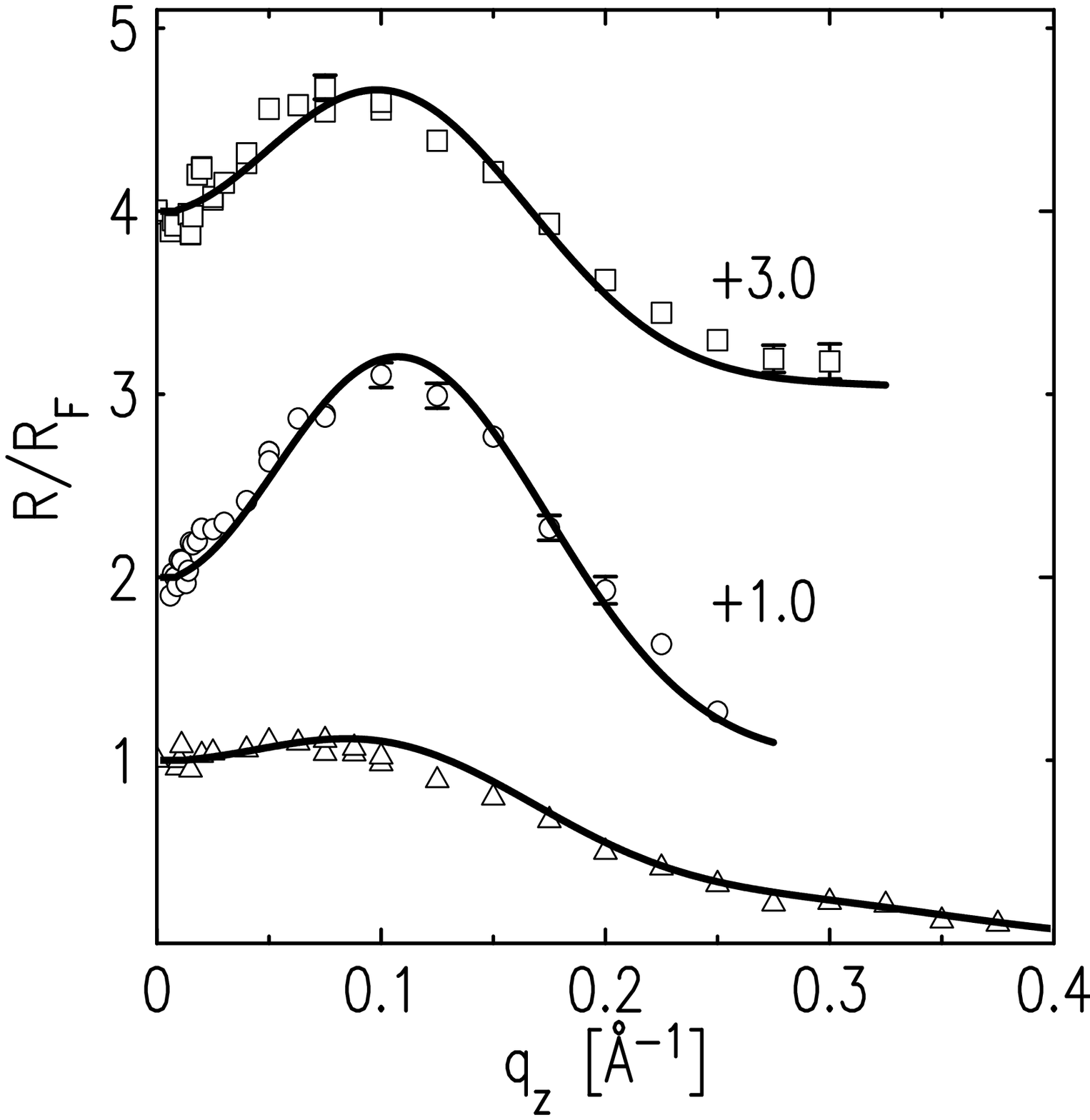, width=0.45\textwidth}

\small {\bf Figure 4.} \it Normalized reflection coefficient $R/R_F$ as a function of $q_z$ for interfaces with an adsorbed octadecanoic acid film: (triangles) n-hexane - water interface at $T = 295$\,K; (circles, squares) toluene - water interface at $T = 308$ and 328\,K, respectively. Solid lines are calculations with qualitative model (2). The numerals at the curves indicate their shifts along the coordinate axis for convenience of presentation.

\end{figure}

According to the fitting of the experimental data with Eq. (4), the electron concentration in the Gibbs
monolayer at the n-hexane–water interface is $\rho_1 = 0.34 \pm 0.01$\,{\it e$^-$/}{\AA}$^3$ and its thickness is $z_1=26 \pm 1$\,{\AA}. These values correspond to $A=\Gamma/(z_1\rho_1)$$=18\pm2$\,\AA$^2$ area per molecule, where $\Gamma=160$ is the number of electrons in the C$_{18}$H$_{36}$O$_2$ molecule. The calculated total length of the octadecanoic acid molecule is $L \approx 25.6$\,\AA{}($=17\times 1.27$\,\AA (Ñ-Ñ) + 1.5\,\AA (-ÑÍ$_3$) + 2.5\,\AA (-ÑÎÎÍ)). Thus, a solid phase of the C$_{18}$-acid monolayer with aliphatic -C$_{17}$H$_{35}$ tails, which are fully ordered and extended along the normal to the surface,
exists at the n-hexane - water interface.

The electron concentration in the Gibbs monolayer at the toluene–water interface at $T=308$\,K is
$\rho_1 = 0.35 \pm 0.01$\,{\it e$^-$/}{\AA}$^3$ and $z_1=22 \pm 1$\,{\AA}. For these parameters, we have $A=\Gamma/(z_1\rho_1)$$=20\pm2$\,\AA$^2$, which also corresponds to the solid phase of the monolayer but with tilted aliphatic tails (deviated by $\theta = \arccos(z_1/L)\approx 30^\circ$ from the normal). Finally, at $T=328$\,K, we have $\rho_1 = 0.26 \pm 0.01$\,{\it e$^-$/}{\AA}$^3$, $z_1=22 \pm 1$\,{\AA},
and $A=24\pm2$\,\AA$^2$. This phase can be conventionally called liquid, since its density corresponds to that of liquid n-octadecane C$_{18}$H$_{38}$ [36].

The values of parameters $A$ and $\theta=0$ of the soluble octadecanoic acid Gibbs monolayer on the n-hexane - water interface are close to the characteristics of the untilted solid phases of the insoluble octadecanoic acid Langmuir monolayer on the water surface [43, 44]. These phases can be represented by either hexagonal phase $LS$ or distorted hexagonal phase $S$ [45]. The parameters of the Gibbs monolayer at the interface with toluene correspond to the characteristics of the tilted hexatic $L_{2d}$ and $Ov$ solid phases of the Langmuir C$_18$-acid monolayer [46]. Moreover, the structure of the soluble monolayer phases is assumed to be similar to the molecular packing in the bulk crystalline rotator phases ($R_I$, $R_{II}$) of a high-molecular-weight saturated hydrocarbon near its melting temperature [47-49].

Thus, octadecanoic acid molecules form an ordered solid monolayer on both the n-hexane - water
and toluene - water interfaces. The microscopic mechanism of forming the solid monolayer is likely to be
based on the formation of a two-dimensional network of hydrogen bonds between the carbonyl (C=O) and
hydroxyl (–OH) groups of neighboring molecules [50]. Our experimental data also illustrate a solid - liquid
phase transition in the Gibbs monolayer on the toluene–water interface, which is accompanied by the
disordering of -C$_{17}$H$_{35}$ hydrocarbon tails. As temperature increases in a certain vicinity of $T_c$, some adsorbed C$_{18}$-acid molecules leave the interface and dissolve in the toluene volume. In this case, $A$ increases by 10–20\% and the detected change in the monolayer thickness $z_1$ is insignificant.

The use of toluene as the upper phase for structural studies has both advantages and disadvantages. On the
one hand, the difference between the volume electron concentrations at the toluene - water interface ($\Delta\rho_1=\rho_w - \rho_t \approx 0.05$\,$e^-$/\AA$^3$) is noticeably lower than that at the hexane - water interface ($\Delta\rho_2=\rho_w - \rho_h \approx 0.10$\,$e^-$/\AA$^3$); therefore, the structure factor oscillation amplitude for the former interface is higher than that for the latter by a factor of $(\Delta\rho_2/\Delta\rho_1)^2 \approx 4$ due to the fact that $R/R_F \propto  \Delta\rho^{-2}$ according to Eq. (4). This fact explains the higher $R/R_F$ oscillation for the toluene - water interface as compared to that for the n-hexane - water interface in Fig. 3.

On the other hand, the interfacial tension of the toluene - water interface is lower than that of the n-hexane - water interface by a factor of 2–3. Therefore, the intensity of nonspecular (diffuse) scattering by the capillary wave roughness of the former interface is substantially higher than that of the latter. This fact and the smaller depth of penetration of photons with $E\approx15$\,keV into toluene as compared to that into n-hexane (18 and 24 cm, respectively) substantially (by about 30\%) decrease the grazing angle range to be measured and to determine $q_z^{max}$ in Eq. (3).

Thermotropic phase transitions, which usually have a desorption origin, were detected in the layers of
high-molecular fatty acids adsorbed on, e.g., the n-hexane - water interface [5, 15]. These are solid - gas
monolayer and liquid - gas monolayer transitions, where almost all adsorbed lipid molecules leave the
interface and dissolve in the oil volume when temperature increases [10, 51, 52]. Multilayers were found to
exist in the normal alkanol films adsorbed on the neutral n-hexane – water and n-hexadecane - water interfaces when the surfactant hydrocarbon chain length exceeds the solvent molecule chain length by approximately six carbon atoms ($\sim 7$\,\AA) [18, 53]. Later [12, 54-56], thermotropic two-dimensional solid - liquid transitions were detected in films on the saturated hydrocarbon (n-alkane) - water interface. These transitions are often described by two critical temperatures. For example, as follows from the data on scattering by the n-hexane - water interface in the case of protonated (pH=2) triacontanic acid monolayers, the two - dimensional interface crystallization transition at $T_c$ is preceded by transition to multilayer adsorption at $T^*>T_c$ when temperature $T$ increases [19].

\begin{figure}
\vspace{0.5in}
\hspace{0.15in}
\epsfig{file=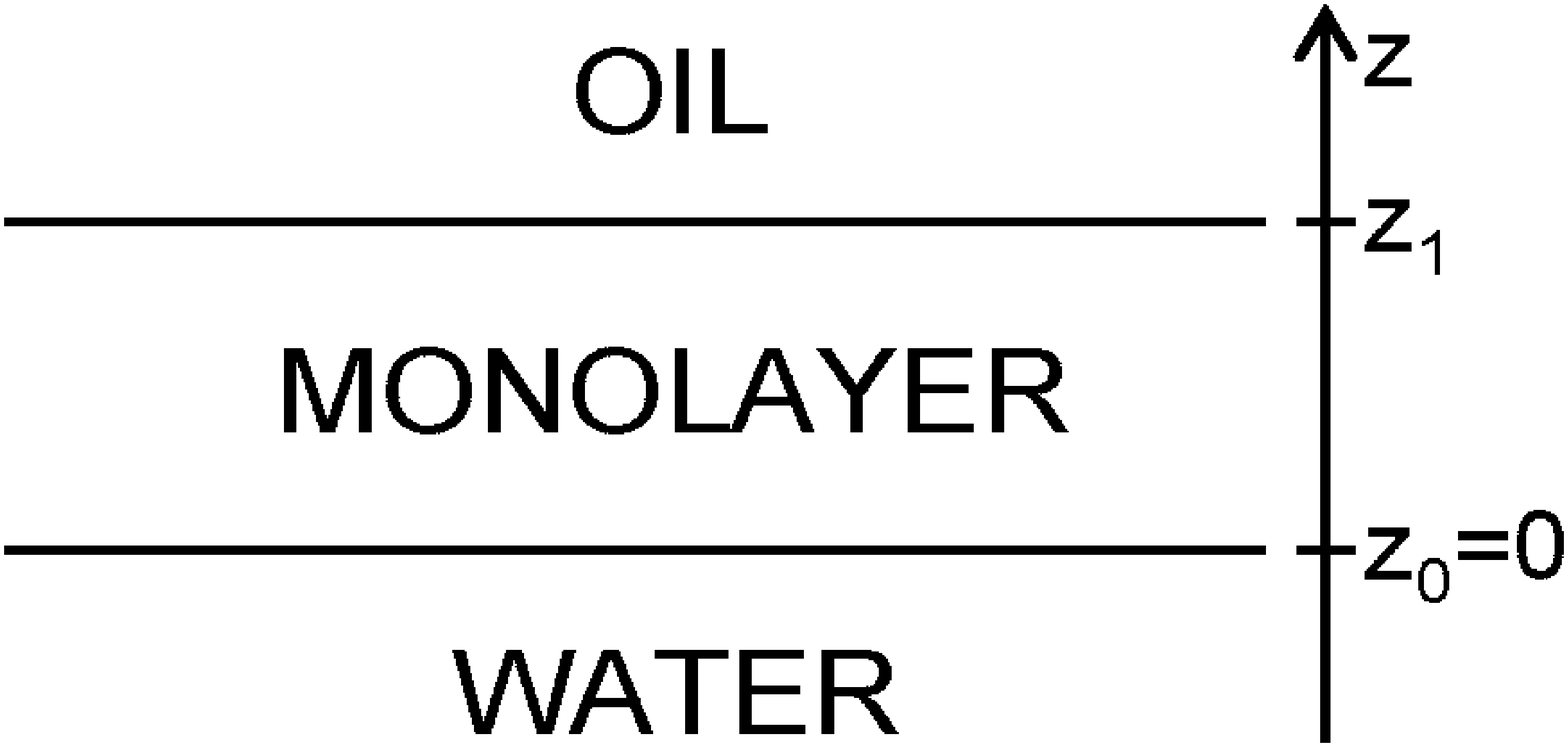, width=0.45\textwidth}

\vspace{0.15in}
\small {\bf Figure 5.} \it Model of the interface.
\end{figure}

\begin{figure}
\epsfig{file=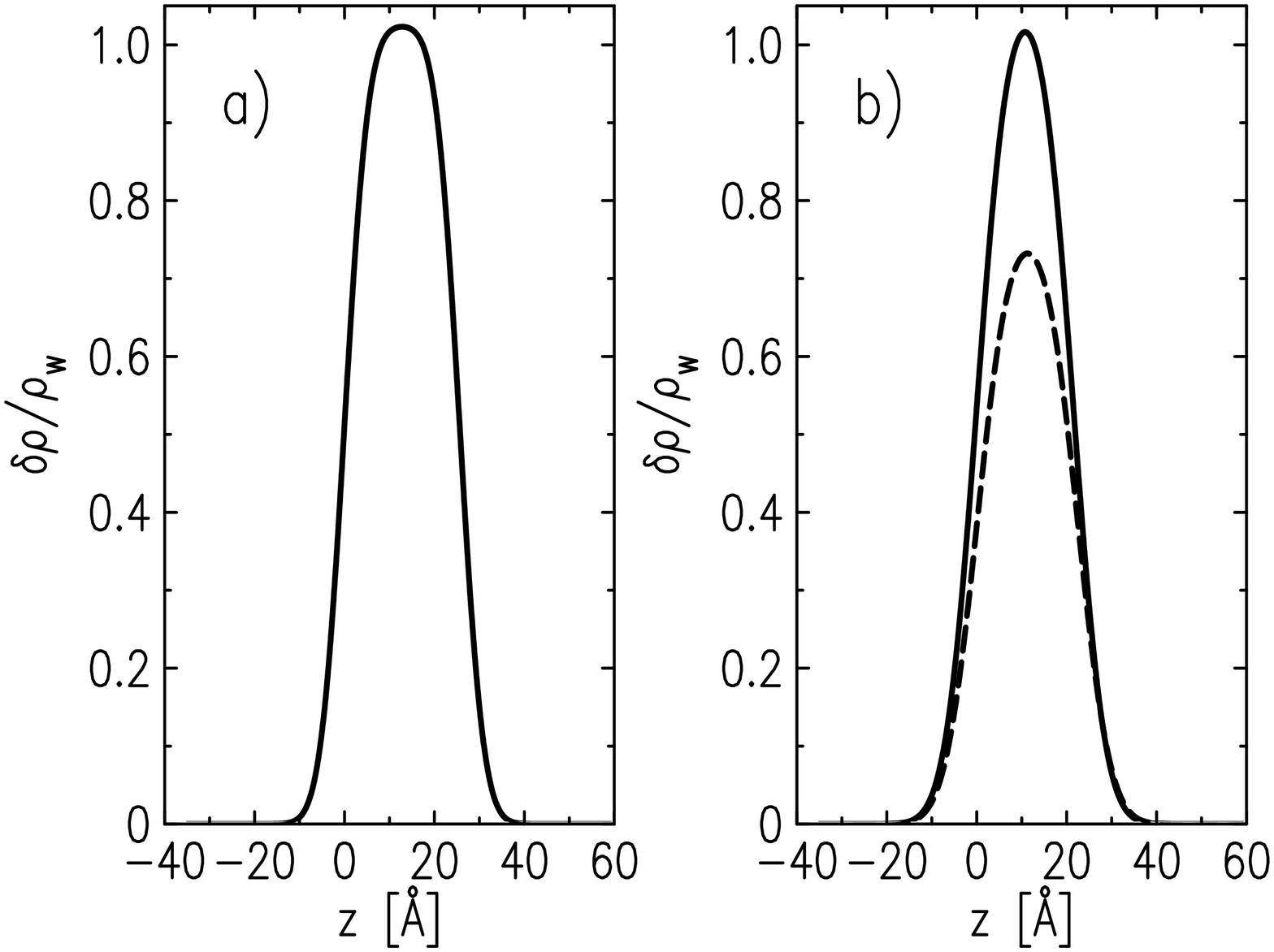, width=0.45\textwidth}

\small {\bf Figure 6.} \it Model electron concentration profiles for octadecanoic
acid monolayer $\delta\rho(z)$ normalized by the electron concentration in water under normal conditions ($\rho_w=0.333$\,{\it e$^-$/}{\AA}$^3$): (a) n-hexane - water interface and (b) toluene - water interface. Model of a solid monolayer phase at $T \approx 308$\,K (solid line) and  model of a liquid monolayer phase at $T\approx328$\,K (dashed line).

\end{figure}

Note that the noncapillary wavy width of the toluene - water interface ($\sigma_{0}\approx 4$\,\AA{}) does not manifest itself in the monolayers adsorbed onto it [57]. As in the case of neutral alkanol monolayers on the n-hexane - water interface (where $\sigma_{0}$ is of the same order of magnitude), we attribute this finding to relatively low contrast in the interfacial structure [18].

As follows from our experimental data, the joint application of reflectometry and diffuse scattering to
the system under study in the vicinity of $T_c$ can give additional useful information about both the possibility of transition to multilayer adsorption of octadecanoic acid at the toluene - water interface and the integrated characteristic of the roughness spectrum ($\sigma_{0}$).

Thus, the phenomena that occur at the interfaces in water - oil emulsions in the presence of impurity
surfactants influence the efficiency of oil technological processes [58-60]. This investigation of the aromatic hydrocarbon - water interface and our earlier studies of the phase transitions at the saturated hydrocarbon - water interface demonstrate fundamentally new experimental abilities for revealing the essence of the processes that occur in oil dispersed systems by X-ray reflectometry and diffuse scattering using synchrotron radiation.

\vspace{0.25in}
{\bf ACKNOWLEDGMENTS}

This work was performed using the resources of the National Synchrotron Light Source, US Department
of Energy (DOE) Office of Science User Facility, operated for the DOE Office of Science by the
Brookhaven National Laboratory under contract no. DE-AC02-98CH10886. The X19C beamline was supported
by the ChemMatCARS National Synchrotron Resource, University of Chicago, University of Illinois
at Chicago, and Stony Brook University. The theoretical part of the work was supported by the Russian Science Foundation (project no. 18-12-00108).

\end{document}